\documentstyle[12pt]{article}
\begin{document}  \bibliographystyle{unsrt}

\begin{flushleft}
Presented at the Fourth International Wigner Symposium\\
(Guadalajara, Mexico, August 7 - 11, 1995).
\end{flushleft}

\vspace{20mm}

\begin{center}

{\large \bf Lorentz Boosts as Squeeze Transformations \\[2mm]
and Coherence Problems}

D. Han\\
{\bf National Aeronautics and Space Administration,\\
Goddard Space Flight Center, Code 910.1,\\
Greenbelt, Maryland 20771, U.S.A.}

Y. S. Kim\\
{\bf Department of Physics, University of Maryland, \\
College Park, Maryland 29742, U.S.A.}

\end{center}

\begin{abstract}
The quark model and the parton model are known to be two different
manifestations of the same covariant entity.  However, the interaction
amplitudes of partons are incoherent while they are coherent in
the quark model.  According to Feynman, this is due to the dilation of
the interaction time among the quarks.  We present a quantitative
analysis of this time-dilation problem using Lorentz boosts as squeeze
transformations.
\end{abstract}

Hadrons are believed to be quantum bound states of quarks having
localized probability distribution.
The most convincing evidence for this bound-state picture is
the hadronic mass spectra which are observed in high-energy laboratories
\cite{fkr71,knp86}.  However, this picture of bound states is applicable
only to observers in the Lorentz frame in which the hadron is at rest.
How would the hadrons appear to observers in other Lorentz frames?

In 1969, Feynman observed that a fast-moving hadron can be regarded
as a collection of many ``partons'' whose properties appear to be quite
different from those of quarks \cite{fey69}.  He made the following
systematic observations.

\begin{itemize}

\item[a] The picture is valid only for hadrons moving with velocity close
to that of light.

\item[b] The interaction time between the quarks becomes dilated, and
partons appear to be dilated, and the partons behave as free particles.
\item[c] The momentum distribution of partons becomes widespread as the
hadron.

\item[d] The number of partons seems to be infinite or much larger than
that of the quarks.

\end{itemize}

\noindent  Since Feynman's invention of the parton model, one of the most
challenging problems in particle theory has been how to reconcile the
above observations with the quark model for hadrons at rest, and there is
a model in which the quark model and the parton model are two different
manifestation of one covariant entity, as $E = p^{2}/2m$ and $E = cp$ are
two different manifestations of Einstein's
$E = mc^{2}$.~\cite{knp86,hussar81}.

In this report, we would like to discuss in detail item b, which is about
one of the basic issues in quantum mechanics.  When an external signal
interacts with the quarks inside the hadron, we have to make a coherent
sum of interaction amplitudes before calculating the total cross section
for the hadron.  On the other hand, in the parton model, we calculate the
cross section for each parton first.  The total cross section is therefore
an incoherent sum of parton cross sections.

We believe in covariance, and we believe also in quantum mechanics based
on the superposition principle.  Then does the parton model indicate that
Lorentz boosts destroy superposition principle?  Indeed, it was Feynman
who faced this problem first.  His suggestion was that the interaction
time between the partons is dilated as the hadron moves fast compared
with the time needed for each parton to interact with the external signal.
In this report, we give a quantitative analysis of this problem starting
from the hadronic wave function at rest, using the squeeze property of
the covariant harmonic oscillator formalism.

The Lorentz boost along the z direction is performed according to
\begin{equation}
\pmatrix{z' \cr t'} = \pmatrix{\cosh\eta & \sinh\eta \cr \sinh\eta &
\cosh\eta} \pmatrix{z \cr t} ,
\end{equation}
where $\eta $ is the boost parameter and is $\tanh^{-1}(v/c)$.
In terms of the light-cone variables:
\begin{equation}
u = (z + t)/\sqrt{2} , \qquad v = (z - t)/\sqrt{2} .
\end{equation}
For convenience, we shall call $u$ and $v$ positive and negative
light-cone axes respectively.  They are perpendicular to each other
The transformation takes a much simpler form:
\begin{equation}\label{lorensq}
\pmatrix{u' \cr v'} = \pmatrix{e^{\eta} & 0 \cr 0 & e^{-\eta}}
\pmatrix{u \cr v} .
\end{equation}
If the system is boosted, one axis expands while the other contracts,
as is shown in Fig.~\ref{squeeze}.
This boost is clearly a squeeze transformation.~\cite{hhk89}  The
product of the two light-cone variables is
\begin{equation}
u v = (z^{2} - t^{2})/2 ,
\end{equation}
which is a Lorentz-invariant quantity.

\begin{figure}
\vspace{70mm}
\caption{Lorentz-squeezed hadronic distribution. The major axis of this
ellipse corresponds to the interaction time among the quarks, while the
minor axis measures the time external signal spends inside the hadron.}
\label{squeeze}
\end{figure}

Next, let us consider a hadron consisting of two quarks.  If the
space-time position of two quarks are specified by $x_{a}$ and $x_{b}$
respectively, the system can be described by the variables
\begin{equation}
X = (x_{a} + x_{b})/2 , \qquad x = (x_{a} - x_{b})/2\sqrt{2} .
\end{equation}
The four-vector $X$ specifies where the hadron is in space and
time, while the variable $x$ measures the space-time separation between
the quarks~ \cite{fkr71}.  The simplest wave function in the covariant
oscillator system is~\cite{knp86}
\begin{equation}\label{wf1}
\psi_{0}(z,t) = \left({1\over\pi} \right)^{1/2}
\exp\left\{-{1\over 2} (z^{2} + t^{2}) \right\} ,
\end{equation}
for the hadron at rest.  Quantum excitations along the $z$ direction are
allowed, but there are no time-like oscillations along the $t$ direction.
In terms of the light-cone variables, the wave function takes the form
\begin{equation}\label{wf2}
\psi_{0}(z,t) = \left({1 \over \pi} \right)^{1/2}
\exp \left\{-{1\over 2} (u^{2} + v^{2}) \right\} . 
\end{equation}
If the system is boosted, the wave function becomes~\cite{knp86}
\begin{equation}\label{wf3}
\psi_{\eta }(z,t) = \left({1 \over \pi }\right)^{1/2}
\exp\left\{-{1\over 2}\left(e^{-2\eta }u^{2} +
e^{2\eta }v^{2}\right)\right\} .
\end{equation}
The transition from Eq.(\ref{wf2}) to Eq.(\ref{wf3}) is a squeeze
transformation.  The wave function of Eq.(\ref{wf2}) is distributed
within a circular region in the $u v$ plane, and thus in the $z t$
plane.  On the other hand, the wave function of Eq.(\ref{wf3}) is
distributed in an elliptic region, as in Fig.~\ref{squeeze}.

As the hadronic speed becomes very close to the speed of light, the
elliptic region becomes concentrated along the positive light-cone
axis $u$.  The major axis of the ellipse becomes very large, and the
minor axis along the $v$ axis becomes very small.  They become multiplied
by the factors $\exp{(\eta)}$ and $\eta{(-\eta)}$ respectively.  If the
quarks are confined to the thin elliptic region, they move with the speed
of light, and the major axis of the ellipse represents the interaction
time between the quarks which can now be called partons.  Thus the
interaction time between the quarks becomes dilated by $\exp{(\eta)}$.

While the hadron moves along the positive light-cone axis, the
interaction signal should come from the opposite direction.  It should
then come along the negative light-cone axis.  The interaction time is
approximately the time the signal overlaps with the hadronic distribution.
It becomes contracted by $\exp{(-\eta)}$.  The ratio of the contracted
time to the dilated time is therefore $e^{-2\eta}$.  The energy of each
proton coming out of the Fermilab accelerator is $900 GeV$.  This leads
the ratio to $10^{-6}$.  This is indeed a small number.  We know what
happened one year ago, but do not know what happened one million years
ago.  The external signal is not able to sense the interaction of the
quarks among themselves inside the hadron.

\end{document}